\documentclass[preprint,prd,aps,showpacs,showkeys,nofootinbib]{revtex4}
\usepackage{graphicx}
\usepackage{makecell}
\usepackage{amssymb}
\usepackage{color} 

\usepackage{bm}
\usepackage{url}
\usepackage{float}
\usepackage{subfigure}

\begin{document}
	\title{The Higgs boson decay $h \rightarrow bs$ in the NB-LSSM}
	\author{Cai Guo$^{1,2,3}$,
		Xing-Xing Dong$^{1,2,3,4}$\footnote{dongxx@hbu.edu.cn}, Zhan Cao$^{1,2,3}$,\\Song Gao$^{1,2,3}$, Shu-Min Zhao$^{1,2,3}$\footnote{zhaosm@hbu.edu.cn},
		Tai-Fu Feng$^{1,2,3}$\footnote{fengtf@hbu.edu.cn}}
	\affiliation{
		$^1$ College of Physics Science and Technology, Hebei University, Baoding, 071002, China\\
		$^2$ Hebei Key Laboratory of High-precision Computation and Application of Quantum Field Theory, Baoding, 071002, China\\
		$^3$ Hebei Research Center of the Basic Discipline for Computational Physics, Baoding, 071002, China\\
		$^4$Departamento de F\'{i}sica and CFTP, Instituto Superior T\'{e}cnico, Universidade de Lisboa, Av. Rovisco Pais 1, 1049-001 Lisboa, Portugal}
\begin{abstract}
Within the framework of the next to minimum B-L supersymmetric model (NB-LSSM), we investigate the flavor transition process $\bar B\rightarrow X_s\gamma$. Building upon this foundation, we further discuss the Higgs decay process $h \to bs$ under the constraint from the $\bar B\rightarrow X_s\gamma$ process. Our study reveals that the branching ratio of $h \to bs$ can significantly deviate from the Standard Model (SM) expectation, depending on the values of the new parameters introduced in the model. This finding highlights the modulation of new physics parameters on the Higgs flavor-violating decay and provides important theoretical grounds for exploring new physics beyond the SM through flavor observables.

\end{abstract}

\pacs{12.60.Jv, 12.15.Lk, 13.35.-r}
\keywords{Flavor transition, Branching ratio, Beyond Standard Model}
\maketitle

\section{Introduction}
The discovery of the long-awaited Higgs boson in 2012 by the ATLAS and CMS collaborations at the LHC marks a milestone in particle physics~\cite{ATLAS:2012yve,Ranjan:2014kya}. The measured mass of this SM-like Higgs boson is $m_{h}=125.20\pm0.11 \rm {GeV}$~\cite{ATLAS:2023oaq}. While the SM of particle physics has achieved remarkable success, the discovery of the Higgs boson has also opened a new window for exploring the mechanism of electroweak symmetry breaking and new physics beyond the SM. The decay of the Higgs boson to a bottom quark and a strange antiquark, $h \to bs$, constitutes a highly sensitive probe of flavor-violating interactions beyond the Standard Model (BSM). In the SM, this process is strictly forbidden at tree level and can only proceed via highly suppressed loop diagrams, primarily due to the Glashow--Iliopoulos--Maiani (GIM) mechanism and the small off-diagonal elements of the Cabibbo--Kobayashi--Maskawa (CKM) matrix. Consequently, the branching ratio $\mathrm{Br}(h \to bs)$ is extremely small, with SM predictions in the range of approximately $10^{-8}$ to $10^{-7}$~\cite{Benitez-Guzman:2015ana,Aranda:2020tqw,Blankenburg:2012ex,Liu:2020nsm}, rendering it experimentally inaccessible at present. This pronounced suppression renders $ h \to bs  $ a unique and powerful probe for new physics. Any significant deviation from the SM expectation would provide clear evidence for new physics beyond the SM.

Numerous BSM extensions can substantially enhance flavor-violating couplings in the Higgs--quark sector. A systematic study within an effective field theory framework was performed in Ref.~\cite{QFV}, where all tree-level ultraviolet completions generating dimension-six operators relevant to flavor-violating Higgs--quark interactions were comprehensively enumerated. These simplified models primarily fall into two categories: those incorporating vector-like quarks and those featuring additional scalar degrees of freedom. Low-energy flavor observables---in particular meson mixing amplitudes and the radiative decay $\bar{B} \to X_s \gamma$---impose stringent constraints on the Wilson coefficients of these operators. In scenarios with vector-like quarks, flavor-violating transitions are generically strongly suppressed, resulting in $\mathrm{Br}(h \to bs)$ far below current experimental sensitivity~\cite{LFV11,LFV22}. In contrast, general two-Higgs-doublet models (2HDMs), particularly those without discrete symmetries enforcing natural flavor conservation, permit sizable flavor-violating Yukawa couplings, thereby allowing considerably larger decay rates~\cite{LFV33,LFV44}. These enhancements are especially prominent in the alignment limit or in parameter regions compatible with the observed Higgs properties. Furthermore, additional particle states (such as extra Higgs doublets, supersymmetric partners, vector-like quarks, etc.) can participate in the loop diagrams, effectively disrupting the delicate GIM cancellation mechanism inherent to the SM. This disruption can potentially elevate the branching ratio of $h \to bs$ by several orders of magnitude, bringing it within reach of current or near-future experimental detection~\cite{Arco:2023hmz}. Similarly, recent studies in extended models such as the $U(1)_X$SSM indicate that $\mathrm{Br}(h \to bs)$ can deviate significantly from the SM prediction, with the magnitude of this deviation being highly sensitive to the values of the newly introduced parameters~\cite{Gao:2024uvd}. They perform comprehensive numerical simulations, taking into account all relevant theoretical and phenomenological constraints. The results show that Br($h\rightarrow bs$) are still allowed at the subpercent (percent) level, which are being explored at the LHC.

\emph{B} physics constitutes one of the most promising avenues for discovering new physics beyond the Standard Model, owing to the relatively small impact of non-perturbative QCD uncertainties on the theoretical predictions of key observables. Recently, the average experimental data on the branching ratios of $\bar B\rightarrow X_s\gamma$ is shown as~\cite{HFLAV:2019otj,ParticleDataGroup:2018ovx}
\begin{eqnarray}
	&&\mathrm{Br}(\bar B\rightarrow X_s \gamma)=(3.40\pm0.21)\times 10^{-4}.
	\label{experimental data}
\end{eqnarray}
The SM predicts the $\bar B\rightarrow X_s\gamma$ branching ratios to be~\cite{Misiak:2020vlo,Misiak:2006zs,Szydagis:2020isq}
\begin{eqnarray}
	&&\mathrm{Br}(\bar B\rightarrow X_s \gamma)=(3.36\pm0.23)\times 10^{-4}
\end{eqnarray}
which are in agreement with the experimental results very
well. So, the precise measurements on the rare \emph{B}-decay processes constrain the new physics beyond SM strictly. It is noteworthy that both $h \to bs$ and $\bar B\rightarrow X_s\gamma$ originate from the $b \to s$ flavor-changing neutral current (FCNC) transition. In new physics frameworks such as the 2HDMs or supersymmetric models, the new particles that contribute to the $h \to bs$ decay, for example charged Higgs bosons or supersymmetric partners, also enter the loop diagrams responsible for the $\bar B \to X_s\gamma$ process. Consequently, any new physics mechanism that attempts to significantly enhance the $h \to bs$ signal strength must first satisfy the stringent experimental bounds from $\bar{B} \to X_s \gamma$, ensuring its predictions are consistent with the existing data. This intrinsic correlation allows us to use the precise results from $B$ physics experiments to effectively constrain the possible parameter space for Higgs flavor-violating couplings, thereby providing a clear theoretical guidance for the experimental search for rare Higgs decays.


A welcoming extension of the SM is NB-LSSM~\cite{Ahmed2021,Han2025,Barger2009}. Based on the minimum supersymmetric Standard Model (MSSM) ~\cite{Haber1985,Rosiek1990,Feng2009}, NB-LSSM extends the gauge symmetry group to \( SU(3)_C \otimes SU(2)_L \otimes U(1)_Y \otimes U(1)_{B-L} \), where \emph{B} represents the baryon number and \emph{L} stands for the lepton number. The invariance under \( U(1)_{B-L} \) gauge groups imposes R-parity conservation in the MSSM, which prevents proton decay~\cite{Aulakh1999}. The singlet scalar \( S \) can obtain a vacuum expectation value (VEV) \( \langle S \rangle = \frac{v_S}{\sqrt{2}} \sim \text{TeV} \) after breaking the local gauge symmetry, which is motivated to explain the \( \mu \) problem naturally. Through the additional singlet Higgs states and right-handed (s)neutrinos, additional parameter space in the NB-LSSM is released from the LEP, Tevatron and LHC constraints to alleviate the hierarchy problem of the MSSM~\cite{Abdallah2017,Yang2020}. Besides, the NB-LSSM can also provide much more DM candidates~\cite{Khalil2009,Basso2012,Rose2017,Rose2018}. Within this theoretical framework, a systematic study of $\bar{B} \to X_s \gamma$ and $h \to bs$ processes carries profound significance. On the one hand, the rich set of new physics parameters introduced by the NB-LSSM, such as the $U(1)_{B-L}$ gauge coupling $g_{B}$, extended Higgs potential parameters, the singlet vacuum expectation value $\langle S \rangle$, and soft SUSY-breaking parameters, which generate substantial new physics corrections to Wilson coefficients in the effective Hamiltonian through loop diagrams.
On the other hand, these corrections manifest in high-precision $B$-physics observables, such as decay branching ratios, thereby providing an exceptionally sensitive low-energy probe for testing the model.

The structure of this paper is organized as follows. In Section~II, we present the ingredients of the NB-LSSM by introducing its superpotential, the general soft breaking terms, new corrected mass matrices and couplings. In Section~III, we investigate the $\bar{B} \to X_s \gamma$ process and then study the $h \to bs$ decay in the mass eigenstate basis. Section~IV is devoted to the presentation and analysis of the numerical results. Finally, the conclusions are provided in Section~V. Some analytical formulas are in the Appendix.

\section{introduction of the NB-LSSM}
Using the local gauge group $U(1)_{B-L}$, we extend the MSSM to obtain the NB-LSSM with the local gauge group $SU(3)_C\times SU(2)_L \times U(1)_Y\times U(1)_{B-L}$. Because of the introduction of three Higgs singlets, the Higgs mass squared matrix is $5\times5$. This can not only explains the 125GeV Higgs mass easily, but also enriches the Higgs physics.
\begin{table}[t]
	\caption{ The chiral superfields and quantum numbers in NB-LSSM}
	\begin{tabular}{|c|c|c|c|c|}
		\hline
		Superfields & $U(1)_Y$ & $SU(2)_L$ & $SU(3)_C$ & $U(1)_{B-L}$ \\
		\hline
		$\hat{q}$ & 1/6 & 2 & 3 & 1/6  \\
		\hline
		$\hat{l}$ & -1/2 & 2 & 1 & -1/2  \\
		\hline
		$\hat{H}_d$ & -1/2 & 2 & 1 & 0 \\
		\hline
		$\hat{H}_u$ & 1/2 & 2 & 1 & 0 \\
		\hline
		$\hat{d}$ & 1/3 & 1 & $\bar{3}$ & -1/6  \\
		\hline
		$\hat{u}$ & -2/3 & 1 & $\bar{3}$ & -1/6 \\
		\hline
		$\hat{e}$ & 1 & 1 & 1 & $1/2$  \\
		\hline
		$\hat{\nu}$ & 0 & 1 & 1 & $1/2$ \\
		\hline
		$\hat{\chi}_1$ & 0 & 1 & 1 & -1 \\
		\hline
		$\hat{\chi}_2$ & 0 & 1 & 1 & 1\\
		\hline
		$\hat{S}$ & 0 & 1 & 1 & 0 \\
		\hline
	\end{tabular}
	\label{quarks}
\end{table}

In Table~\ref{quarks}, we show the chiral superfields and quantum numbers of the NB-LSSM. The corresponding superpotential of the NB-LSSM is shown as
\begin{eqnarray}
	&&W=-Y_d\hat{d}\hat{q}\hat{H}_d-Y_e\hat{e}\hat{l}\hat{H}_d-\lambda_2\hat{S}\hat{\chi}_1\hat{\chi}_2+\lambda\hat{S}\hat{H}_u\hat{H}_d+\frac{\kappa}{3}\hat{S}\hat{S}\hat{S}+Y_u\hat{u}\hat{q}\hat{H}_u+Y_{\chi}\hat{\nu}\hat{\chi}_1\hat{\nu}
	\nonumber\\&&~~~~~~~+Y_\nu\hat{\nu}\hat{l}\hat{H}_u.
\end{eqnarray}
Here, $\hat{\chi}_1,~\hat{\chi}_2,~\hat{S}$ are three Higgs singlets. $Y_{u,d,e,\nu,\chi}$ are the Yukawa couplings. $\lambda$, $\lambda_2$ and $\kappa$ are the dimensionless couplings. These couplings can influence the FCNC decay $h\rightarrow bs$ through the one-loop contributions. It is important to note that the term \(Y_{\nu}^{\prime} \hat{\nu} \hat{l} \hat{S}\) is not allowed, as the sum of \(U(1)_{Y}\) charges of \(\hat{\nu}, \hat{l}, \hat{S}\) does not satisfy the necessary charge neutrality condition.

The soft SUSY breaking terms are
\begin{eqnarray}
	&&\mathcal{L}_{soft}=\mathcal{L}_{soft}^{MSSM}-\frac{T_\kappa}{3}S^3+T_{\lambda}SH_d^iH_u^j+T_{2}S\chi_1\chi_2\nonumber\\&&
	-T_{\chi,ik}\chi_1\tilde{\nu}_{R,i}^{*}\tilde{\nu}_{R,k}^{*}
	-T_{\nu,ij}H_u^i\tilde{\nu}_{R,i}^{*}\tilde{e}_{L,j}-m_{\eta}^2|\chi_1|^2-m_{\bar{\eta}}^2|\chi_2|^2\nonumber\\&&-m_S^2|S|^2-m_{\nu,ij}^2\tilde{\nu}_{R,i}^{*}\tilde{\nu}_{R,j} -\frac{1}{2}(2M_{BB^\prime}\tilde{B}\tilde{B^\prime}+M_{BL}\tilde{B^\prime}^2+h.c.).
\end{eqnarray}
$\mathcal{L}_{soft}^{MSSM}$ represents the soft breaking term in the MSSM. $T_{\kappa}$, $T_{\lambda}$, $T_2$, $T_{\chi}$ and $T_{\nu}$ are all trilinear coupling coefficients. For the soft breaking up-squark mass matrices $m^{2}_{\tilde{u}, \tilde{q}}$ and the trilinear coupling matrix $T_{u}$, we introduce the up-squark flavor mixings, which take into account the off-diagonal terms~\cite{Arganda2016,Zhang2014,Calibbi2018}. These mixings are parametrized by means of a complete set of up-squark flavor mixing dimensionless parameters $\delta^{XX}_{ij}$ with $XX = LL, RR, LR(RL=LR)$~\cite{Arganda2016}, and flavor indices $i, j = 1, 2, 3$, with $i \neq j$,
\begin{eqnarray}
&&m_{\tilde{q}}^2 = \left(\begin{array}{ccc}
		m_{Q}^2 & \delta_{12}^{LL}m_{qq}^2 &  \delta_{13}^{LL}m_{qq}^2\\
		\delta_{12}^{LL}m_{qq}^2 & m_{Q}^2 & \delta_{23}^{LL}m_{qq}^2\\
		\delta_{13}^{LL}m_{qq}^2 & \delta_{23}^{LL}m_{qq}^2 & m_{Q}^2
\end{array}
\right), \nonumber\\&&
	m_{\tilde{u}}^2 = \left(\begin{array}{ccc}
		m_{U}^2 & \delta_{12}^{RR}m_{uu}^2 &  \delta_{13}^{RR}m_{uu}^2 \\
		\delta_{12}^{RR}m_{uu}^2 & m_{U}^2 & \delta_{23}^{RR}m_{uu}^2 \\
		\delta_{13}^{RR}m_{uu}^2 & \delta_{23}^{RR}m_{uu}^2 & m_{U}^2
	\end{array}
\right), \nonumber\\&&
	T_u = \left(\begin{array}{ccc}
		1 & \delta_{12}^{LR} &  \delta_{13}^{LR} \\
		\delta_{12}^{LR} & 1 & \delta_{23}^{LR} \\
		\delta_{13}^{LR} & \delta_{23}^{LR} & 1
	\end{array}
\right) A_u. \label{eq:transformation_matrix} 
\end{eqnarray}
We show the concrete forms of the two Higgs doublets and three Higgs singlets
\begin{eqnarray}
	&&\hspace{-2cm}H^0_d=\frac{1 }{\sqrt{2}} \phi_{d}+{\frac{1 }{\sqrt{2}} }v_{d}+i{\frac{1 }{\sqrt{2}} }\sigma_d,
	\nonumber\\&&\hspace{-2cm}H^0_u={\frac{1 }{\sqrt{2}} }\phi_{u}+{\frac{1 }{\sqrt{2}} }v_{u}+i{\frac{1 }{\sqrt{2}} }\sigma_u,
	\nonumber\\&&\hspace{-2cm}\chi_1={\frac{1 }{\sqrt{2}} }\phi_{1}+{\frac{1 }{\sqrt{2}} }v_{\eta}+i{\frac{1 }{\sqrt{2}} }\sigma_1,
	\nonumber\\&&\hspace{-2cm}\chi_2={\frac{1 }{\sqrt{2}} }\phi_{2}+{\frac{1 }{\sqrt{2}} }v_{\bar{\eta}}+i{\frac{1 }{\sqrt{2}} }\sigma_2,
	\nonumber\\&&\hspace{-2cm}S={\frac{1 }{\sqrt{2}} }\phi_S+{\frac{1 }{\sqrt{2}} }v_S+i{\frac{1 }{\sqrt{2}} }\sigma_S.
\end{eqnarray}
The VEVs of the Higgs superfields $H_u$, $H_d$, $\chi_1$, $\chi_2$ and $S$ are presented by
$v_u,~v_d,~v_\eta,~v_{\bar\eta}$ and $v_S$ respectively. Two angles are defined as
$\tan\beta=v_u/v_d$ and $\tan\beta^{\prime }=v_{\bar{\eta}}/v_{\eta}$.

$U(1)_Y$ and $U(1)_{B-L}$ have the gauge kinetic mixing, which can also be induced through RGEs even with zero value at $M_{GUT}$. The covariant derivatives of this model can be written as
\begin{eqnarray}
	&&D_\mu=\partial_\mu-i\left(\begin{array}{cc}Y,&B-L\end{array}\right)
	\left(\begin{array}{cc}g_{Y} &g{'}_{{YB}}\\g{'}_{{BY}} &g{'}_{{B-L}}\end{array}\right)
	\left(\begin{array}{c}A_{\mu}^{\prime Y} \\ A_{\mu}^{\prime BL}\end{array}\right)\;,
	\label{gauge1}
\end{eqnarray}
where $Y$ and $B-L$ represent the hypercharge and $B-L$ charge, respectively. The two Abelian gauge groups are unbroken, then the change of basis can occur with the rotation matrix $R$ ($R^TR=1$)~\cite{Belanger2017,Barger2009,Chankowski2006,Yang2018},
\begin{eqnarray}
	&&\left(\begin{array}{cc}g_{Y} &g{'}_{{YB}}\\g{'}_{{BY}} &g{'}_{{B-L}}\end{array}\right)
	R^T=\left(\begin{array}{cc}g_{1} &g_{{YB}}\\0 &g_{{B}}\end{array}\right)\;.
	\label{gauge3}
\end{eqnarray}
As a result, the $U(1)$ gauge fields are redefined as
\begin{eqnarray}
	&&R\left(\begin{array}{c}A_{\mu}^{\prime Y} \\ A_{\mu}^{\prime BL}\end{array}\right)
	=\left(\begin{array}{c}A_{\mu}^{Y} \\ A_{\mu}^{BL}\end{array}\right)\;.
	\label{gauge4}
\end{eqnarray}

In the NB-LSSM, four two-component spinors ($\tilde{W}^{-}$, $\tilde{W}^{+}$, $\tilde{H}^{-}$, $\tilde{H}^{+}$) form two four-component Dirac fermions (Charginos) $\chi^{+}$, $\chi^{-}$
\begin{eqnarray}
	M_{\chi^{\pm}}=\left(
    \begin{array}{cc}
		M_{2} & \frac{1}{\sqrt{2}}g_{2}v_{u} \\
		\frac{1}{\sqrt{2}}g_{2}v_{d} & \frac{1}{\sqrt{2}} \lambda v_S
	\end{array}
\right).
\end{eqnarray}
This matrix is diagonalized by the unitary matrices $U$ and $V$:
\begin{equation}
	U^{*} M_{\chi^{\pm}} V^{\dagger}=M_{\chi^{\pm}}^{\text {diag }}.
\end{equation}

Based on $(\tilde{u}_{L}, \tilde{u}_{R})$, the mass squared matrix for Up-Squarks reads
\begin{eqnarray}
	&&\tilde{M}^{2}_{\tilde{U}}=\left(
\begin{array}{cc}
		(M^{2}_{U})_{LL} &\frac{1}{2} \left( -\lambda v_d v_s Y_u^{\dagger} + \sqrt{2} v_u T_u^{\dagger} \right) \\
		\frac{1}{2} \left( \sqrt{2} v_u T_u - v_d v_s Y_u \lambda^{*} \right) & (M^{2}_{U})_{RR}
	\end{array}
\right),
\end{eqnarray}
\begin{eqnarray}
&&(M^{2}_{U})_{LL} =  \frac{1}{24} \Big[ 3g_2^2 (v_d^2 - v_u^2) + (g_1^2 + g_{YB}^2)(v_u^2 - v_d^2)
	- 2 g_B^2(v_{\bar{\eta}}^2 - v_{\eta}^2) \nonumber \\&&\hspace{2cm}
	 + (g_1+ g_{YB} g_B)(2v_{\eta}^2 - 2v_{\bar{\eta}}^2 - v_d^2 + v_u^2) \Big]+ \frac{1}{2} (2m_{\tilde{q}}^2 + v_u^2 Y_u^{\dagger} Y_u),
	\label{eq111}
\end{eqnarray}
\begin{eqnarray}
&&	(M^{2}_{U})_{RR} =  \frac{1}{24} \Big[ 2g_B^2(v_{\bar{\eta}}^2 - v_{\eta}^2) + 4(g_1^2 + g_{YB}^2)(v_d^2 - v_u^2) \nonumber \\&&\hspace{2cm}
	 + (g_1+ g_{YB} g_B)(-8v_{\eta}^2 + 8v_{\bar{\eta}}^2 - v_u^2 + v_d^2) \Big] + \frac{1}{2} (2m_{\tilde{u}}^2 + v_u^2 Y_u Y_u^{\dagger}) .
	\label{eq112}  
\end{eqnarray}
This matrix is diagonalized by the unitary matrix $Z^{U}$:\begin{eqnarray}
	&&Z^{U} \tilde{M}^{2}_{\tilde{U}} Z^{U,\dagger}=({M^{2}_{\tilde{U}}})^{\text {diag }}.
\end{eqnarray}

The mass matrix for Higgs boson in the basis $(\phi_d,\phi_u,\phi_1,\phi_2,\phi_S) $ is
\begin{equation}
	m_{\tilde{h}} ^2 = \left(
	\begin{array}{cccccccc}
		m_{\phi_d} m_{\phi_d}  &m_{\phi_u} m_{\phi_d} &m_{\phi_1} m_{\phi_d} &m_{\phi_2} m_{\phi_d} &m_{\phi_S} m_{\phi_d}\\
		m_{\phi_d} m_{\phi_u}  &m_{\phi_u} m_{\phi_u} &m_{\phi_1} m_{\phi_u} &m_{\phi_2} m_{\phi_u} &m_{\phi_S} m_{\phi_u}\\
		m_{\phi_d} m_{\phi_1}  &m_{\phi_u} m_{\phi_1} &m_{\phi_1} m_{\phi_1} &m_{\phi_2} m_{\phi_1} &m_{\phi_S} m_{\phi_1}\\
		m_{\phi_d} m_{\phi_2}  &m_{\phi_u} m_{\phi_2} &m_{\phi_1} m_{\phi_2} &m_{\phi_2} m_{\phi_2} &m_{\phi_S} m_{\phi_2}\\
		m_{\phi_d} m_{\phi_S}  &m_{\phi_u} m_{\phi_S} &m_{\phi_1} m_{\phi_S} &m_{\phi_2} m_{\phi_S} &m_{\phi_S} m_{\phi_S}
	\end{array}
	\right).\label{higgs}
\end{equation}
\begin{eqnarray}
&&	m_{\phi_d} m_{\phi_d} = \frac{1}{4} G^2 v_d^2 - \frac{1}{4} \left( -2\sqrt{2} v_S \Re(T_\lambda)+ (-\kappa v_S^2 + \lambda_2 v_{\eta} v_{\bar{\eta}}  )\lambda^* + \lambda(v_{\eta} v_{\bar{\eta}} \lambda_2^* - v_S^2 \kappa^*) \right) \tan\beta \nonumber\\&&
	m_{\phi_d} m_{\phi_u} =-\frac{1}{4} G^2 v_d v_u + \frac{1}{4} \left( -2 \sqrt{2} v_S \Re(T_\lambda) + (4\lambda v_d v_u - \kappa v_S^2 + \lambda_2 v_{\eta} v_{\bar{\eta}})\lambda^* + \lambda (v_{\eta} v_{\bar{\eta}} \lambda_2^* - v_S^2 \kappa^*) \right) \nonumber\\&&
	m_{\phi_u} m_{\phi_u} =\frac{1}{4} G^2 v_u^2 - \frac{1}{4} \left( -2\sqrt{2} v_S \Re(T_\lambda) + (-\kappa v_S^2 + \lambda_2 v_{\eta} v_{\bar{\eta}})\lambda^* + \lambda(v_{\eta} v_{\bar{\eta}} \lambda_2^* - v_S^2 \kappa^*) \right) \cot\beta \nonumber\\&&
	m_{\phi_d} m_{\phi_1} = \frac{1}{2}(g_1 + g_{YB} g_B) v_d v_{\eta} + \frac{1}{2} v_u  v_{\bar{\eta}} \Re(\lambda^* \lambda_2) \nonumber\\&&
	m_{\phi_u} m_{\phi_1} = -\frac{1}{2}(g_1 +g_{YB} g_B) v_u v_{\eta} + \frac{1}{2} v_d v_{\bar{\eta}} \Re(\lambda^* \lambda_2) \nonumber\\&&
	m_{\phi_1} m_{\phi_1} = g_B^2 v_{\eta}^2 - \frac{1}{4} \left( -2\sqrt{2} v_S \Re(T_2)+ (-\kappa v_S^2 + \lambda v_d v_u)\lambda_2^*+ \lambda_2 (v_d v_u \lambda^* - v_S^2 \kappa^*) \right) \tan\beta' \nonumber\\&&
	m_{\phi_d} m_{\phi_2} = -\frac{1}{2}(g_{YB} g_B) v_d v_{\bar{\eta}} + \frac{1}{2} v_u v_{\eta} \Re(\lambda^* \lambda_2) \nonumber\\&&
	m_{\phi_u} m_{\phi_2} = \frac{1}{2}(g_{YB} g_B) v_u v_{\bar{\eta}}+ \frac{1}{2} v_d v_{\eta} \Re(\lambda^* \lambda_2) \nonumber\\&&
	m_{\phi_1} m_{\phi_2} = \frac{1}{4}\left(-2 \sqrt{2} v_S \Re(T_2) +\left(4\lambda_2 v_{\eta} v_{\bar{\eta}}-\kappa v_S^2 + \lambda v_d v_u\right)\lambda_2^* + \lambda_2\left(v_d v_u \lambda^* - v_S^2\kappa^*\right) \right)-g_B^2 v_{\eta} v_{\bar{\eta}} \nonumber\\&&
	m_{\phi_2} m_{\phi_2} = g_B^2 v_{\bar{\eta}}^2 - \frac{1}{4} \left( -2\sqrt{2} v_S \Re(T_2) + (-\kappa v_S^2 + \lambda v_d v_u)\lambda_2^* + \lambda_2(v_d v_u \lambda^* - v_S^2 \kappa^*) \right) \cot\beta' \nonumber\\&&
	m_{\phi_d} m_{\phi_S} = \frac{1}{2} \left( v_S \left( 2 \lambda v_d - \kappa v_u \right) \lambda - v_u \left( \lambda v_S \kappa + \sqrt{2} \Re(T_\lambda) \right) \right)\nonumber\\&&
	m_{\phi_u} m_{\phi_S} = \frac{1}{2} \left( -v_d \left( \lambda v_S \kappa^*+ \sqrt{2}  \Re(T_\lambda) \right) - v_S \left( -2 \lambda v_u + \kappa v_d \right) \lambda^* \right)\nonumber\\&&
	m_{\phi_1} m_{\phi_S} = \frac{1}{2} \left( -v_{\bar{\eta}} \left( \lambda_2 v_S \kappa^*+ \sqrt{2}  \Re(T_2) \right) + v_S \left( 2 \lambda_2 v_{\eta}-\kappa v_{\bar{\eta}} \right) \lambda_2^* \right)\nonumber\\&&
	m_{\phi_2} m_{\phi_S} = \frac{1}{2} \left( -v_{\eta} \left( \lambda_2 v_S \kappa^*+ \sqrt{2}  \Re(T_2) \right) - v_S \left( -2 \lambda_2 v_{\bar{\eta}}+\kappa v_{\eta} \right) \lambda_2^* \right)\nonumber\\&&
	m_{\phi_S} m_{\phi_S} = 2 \kappa^* \kappa v_S^2 + \frac{\sqrt{2}}{2} \Re(T_\kappa) v_S + \frac{\sqrt{2}}{2} \Re(T_\lambda) \frac{v_u v_d}{v_S} + \frac{\sqrt{2}}{2} \Re(T_2) \frac{v_{\eta}v_{\bar{\eta}}} {v_S}
	\label{21}
\end{eqnarray}
This matrix is diagonalized by the rotation matrix $Z^H$,
\begin{equation}
	Z^H m_{\tilde{h}}^2 Z^{H \dagger} = (m_{h_i^0}^2)^{diag}.
\end{equation}
Here, $G^2=g_{1}^2+g_{2}^2+g_{{YB}}^2$. The mass of the SM-like Higgs boson can be obtained after considering the leading-log radiative corrections from stop and top particles.
\begin{eqnarray}
	m_{h}=\sqrt{(m_{h_1^0})^2+\Delta m_h^2},
\end{eqnarray}
where $m_{h_1^0}$ represents the lightest tree-level Higgs boson mass, and the leading-log radiative corrections $\Delta m_h^2$ can be written as~\cite{HiggsC1,HiggsC3}
\begin{eqnarray}
	&&\Delta m_h^2=\frac{3m_t^4}{4 \pi ^2 v^2}\Big[\Big(\tilde{t}+\frac{1}{2}\tilde{X}_t\Big)+\frac{1}{16 \pi ^2}\Big(\frac{3m_t^2}{2v^2}-32\pi\alpha_3\Big)(\tilde{t}^2+\tilde{X}_t\tilde{t})\Big],\nonumber \\&&\tilde{t}=\log\frac{M_S^2}{m_t^2},~~~~\tilde{X}_t=\frac{2\tilde{A}_t^2}{M_S^2}(1-\frac{\tilde{A}_t^2}{12M_S^2}).
\end{eqnarray}
Here, $\alpha_3$ is the strong coupling constant, $M_S=\sqrt{m_{\tilde{t}_1}m_{\tilde{t}_2}}$ with $m_{\tilde{t}_{1,2}}$ are the stop masses, $\tilde{A}_t=A_t-\frac{\lambda v_S }{\sqrt{2}}\cot \beta $ with $A_t=T_{u,33}$ denotes the trilinear Higgs-stops coupling.

\section{The Processes $\bar{B} \to X_s \gamma$ and $h \to bs$ in the NB-LSSM}
In this section, we analyze the branching ratios of the processes $\bar{B} \to X_s \gamma$ and $h \to bs$. The concrete details are discussed as follows.
\subsection{The rare decay $\bar{B}\rightarrow X_s\gamma$ }
The effective Hamilton for the transition $\bar{B}\to X_s \gamma$ at hadronic scale can be written as
\begin{eqnarray}
	&&H_{eff}=-\frac{4G_F}{\sqrt{2}}V_{ts}^\ast V_{tb}\Big[C_1\mathcal{O}^c_1+C_2\mathcal{O}_2^c+\sum_{i=3}^6\mathcal{O}_i+\sum_{i=7}^{10}(C_i\mathcal{O}_i+C'_i\mathcal{O}'_i)\nonumber\\
	&&\qquad\;\quad\;+\sum_{i=S,P}(C_i\mathcal{O}_i+C'_i\mathcal{O}'_i)\Big],
\end{eqnarray}
where $\mathcal{O}_i(i=1, 2,...,10, S, P)$ and $\mathcal{O}'_i(i=7, 8,...,10, S, P)$ are defined as~\cite{Altmannshofer:2008dz,Lin:2009zzh,Yang:2010wh,Goertz:2011nx}
\begin{eqnarray}
	&&{\cal O}_{_1}^u=(\bar{s}_{_L}\gamma_\mu T^au_{_L})(\bar{u}_{_L}\gamma^\mu T^ab_{_L})\;,\;\;
	{\cal O}_{_2}^u=(\bar{s}_{_L}\gamma_\mu u_{_L})(\bar{u}_{_L}\gamma^\mu b_{_L})\;,
	\nonumber\\
	&&{\cal O}_{_3}=(\bar{s}_{_L}\gamma_\mu b_{_L})\sum\limits_q(\bar{q}\gamma^\mu q)\;,\;\;
	{\cal O}_{_4}=(\bar{s}_{_L}\gamma_\mu T^ab_{_L})\sum\limits_q(\bar{q}\gamma^\mu T^aq)\;,
	\nonumber\\
	&&{\cal O}_{_5}=(\bar{s}_{_L}\gamma_\mu\gamma_\nu\gamma_\rho b_{_L})\sum\limits_q(\bar{q}\gamma^\mu
	\gamma^\nu\gamma^\rho q)\;,\;\;
	{\cal O}_{_6}=(\bar{s}_{_L}\gamma_\mu\gamma_\nu\gamma_\rho T^ab_{_L})\sum\limits_q(\bar{q}\gamma^\mu
	\gamma^\nu\gamma^\rho T^aq)\;,
	\nonumber\\
	&&{\cal O}_{_7}={e\over 16\pi^2}m_{_b}(\bar{s}_{_L}\sigma_{_{\mu\nu}}b_{_R})F^{\mu\nu}\;,\;\;
	{\cal O}_{_7}'={e\over 16\pi^2}m_{_b}(\bar{s}_{_R}\sigma_{_{\mu\nu}}b_{_L})F^{\mu\nu}\;,\;\;
	\nonumber\\
	&&{\cal O}_{_8}={g_{_s}\over 16\pi^2}m_{_b}(\bar{s}_{_L}\sigma_{_{\mu\nu}}T^ab_{_R})G^{a,\mu\nu}\;,\;\;
	{\cal O}_{_8}'={g_{_s}\over 16\pi^2}m_{_b}(\bar{s}_{_R}\sigma_{_{\mu\nu}}T^ab_{_L})G^{a,\mu\nu}\;,\;\;
	\nonumber\\
	&&{\cal O}_{_9}={e^2\over g_{_s}^2}(\bar{s}_{_L}\gamma_\mu b_{_L})\bar{l}\gamma^\mu l\;,\;\;
	{\cal O}_{_9}'={e^2\over g_{_s}^2}(\bar{s}_{_R}\gamma_\mu b_{_R})\bar{l}\gamma^\mu l\;,\;\;
	\nonumber\\
	&&{\cal O}_{_{10}}={e^2\over g_{_s}^2}(\bar{s}_{_L}\gamma_\mu b_{_L})\bar{l}\gamma^\mu\gamma_5 l\;,\;\;
	{\cal O}_{_{10}}'={e^2\over g_{_s}^2}(\bar{s}_{_R}\gamma_\mu b_{_R})\bar{l}\gamma^\mu\gamma_5 l\;,\;\;
	\nonumber\\
	&&{\cal O}_{_S}={e^2\over16\pi^2}m_{_b}(\bar{s}_{_L}b_{_R})\bar{l}l\;,\;\;
	{\cal O}_{_S}'={e^2\over16\pi^2}m_{_b}(\bar{s}_{_R}b_{_L})\bar{l}l\;,\;\;\nonumber\\
	&&{\cal O}_{_P}={e^2\over16\pi^2}m_{_b}(\bar{s}_{_L}b_{_R})\bar{l}\gamma_5l\;,\;\;
	{\cal O}_{_P}'={e^2\over16\pi^2}m_{_b}(\bar{s}_{_R}b_{_L})\bar{l}\gamma_5l\;,
	\label{operators}
\end{eqnarray}
where $g_s$ denotes the strong coupling, $F^{\mu\nu}$ are the electromagnetic  field strength tensors, $G^{\mu\nu}$ are the gluon field strength tensors, $T^a\,(a=1,...,8)$ are $SU(3)$ generators.

\begin{figure}[t]
	\centering
	\includegraphics[width=0.9\linewidth]{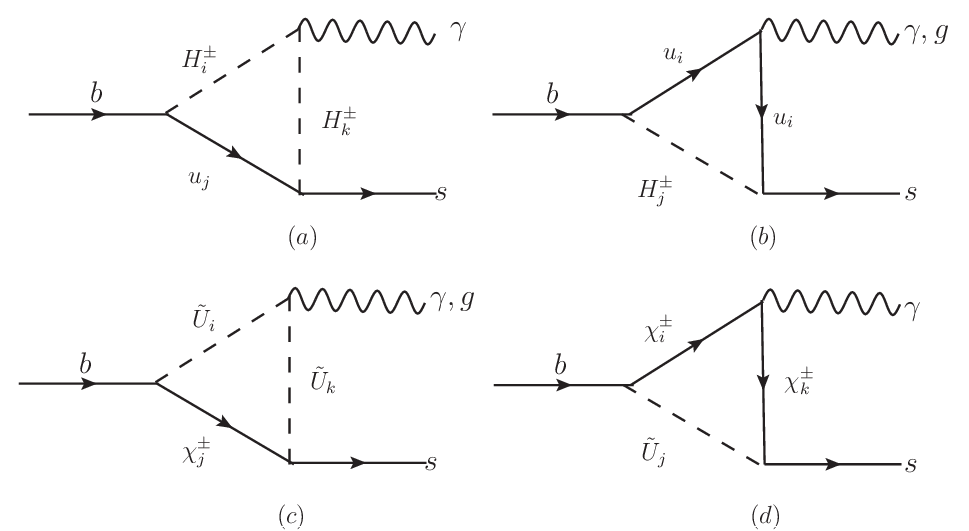}
	\caption[]{The one loop Feynman diagrams contributing to $\bar{B}\rightarrow X_s\gamma$.}
	\label{fig:mia}
\end{figure}

The one-loop Feynman diagrams contributing to the $\bar{B}\rightarrow X_s\gamma$
process are shown in Fig.~\ref{fig:mia}.
Then the branching ratio of $\bar{B}\rightarrow X_s\gamma$ in the NB-LSSM can be written as
\begin{eqnarray}
	&&Br(\bar{B}\rightarrow X_s\gamma)=R\Big(|C_{7\gamma}(\mu_b)|^2
	+N(E_\gamma)\Big)\;,
\end{eqnarray}
where the overall factor $R=2.47\times10^{-3}$, and the nonperturbative contribution
$N(E_\gamma)=(3.6\pm0.6)\times10^{-3}$\cite{Buras:2011zb}. $C_{7\gamma}(\mu_b)$ is defined by
\begin{eqnarray}
	&&C_{7\gamma}(\mu_b)=C_{7\gamma,SM}(\mu_b)+C_{7,NP}(\mu_b),
\end{eqnarray}
where we choose the hadron scale $\mu_b=2.5$ GeV and use the SM contribution at NNLO level $C_{7\gamma,SM}(\mu_b) = -0.3689$~\cite{Goertz:2011nx,Gambino:2001ew}. The Wilson coefficients for new physics at the bottom quark scale can be written as~\cite{Buras:1993xp,Gao:2013fxa}
\begin{eqnarray}
	&&C_{7,NP}(\mu_b)\approx0.5696
	C_{7,NP}(\mu_{EW})+0.1107 C_{8,NP}(\mu_{EW}),
\end{eqnarray}
with
\begin{eqnarray}
	&&C_{7,NP}(\mu_{EW})=C_{7,NP}^{(a)}(\mu_{EW})+C_{7,NP}^{\prime(a)}(\mu_{EW})+C_{7,NP}^{(b)}(\mu_{EW})+C_{7,NP}^{\prime(b)}(\mu_{EW})\nonumber\\&&\qquad\;\qquad\;\qquad\;+C_{7,NP}^{(c)}(\mu_{EW})+C_{7,NP}^{(c)}(\mu_{EW})+C_{7,NP}^{\prime(d)}(\mu_{EW})+
	C_{7,NP}^{\prime(d)}(\mu_{EW})\nonumber\\
	&&C_{8,NP}(\mu_{EW})=C_{8g,NP}(\mu_{EW})+C_{8g,NP}^{\prime}(\mu_{EW}),\nonumber\\
	&&C_{8g,NP}(\mu_{EW})=[C_{7,NP}^{(b)}(\mu_{EW})+C_{7,NP}^{(c)}(\mu_{EW})]/Q_u, \nonumber\\
	&&C_{8g,NP}^{\prime}(\mu_{EW})=[C_{7,NP}^{\prime(b)}(\mu_{EW})+C_{7,NP}^{\prime(c)}(\mu_{EW})]/Q_u,
\end{eqnarray}
where $Q_u=2/3$, $C_{7,NP}(\mu_{EW})$ is Wilson coefficient of the process $b\rightarrow s\gamma$, while $C_{8,NP}(\mu_{EW})$ is Wilson coefficient for the process $b\rightarrow sg$, corresponding to Fig.~\ref{fig:mia}. Their concrete expressions are collected in
Appendix~\ref{bsr}.
\begin{figure}[t]
	\centering
	\includegraphics[width=0.8\linewidth]{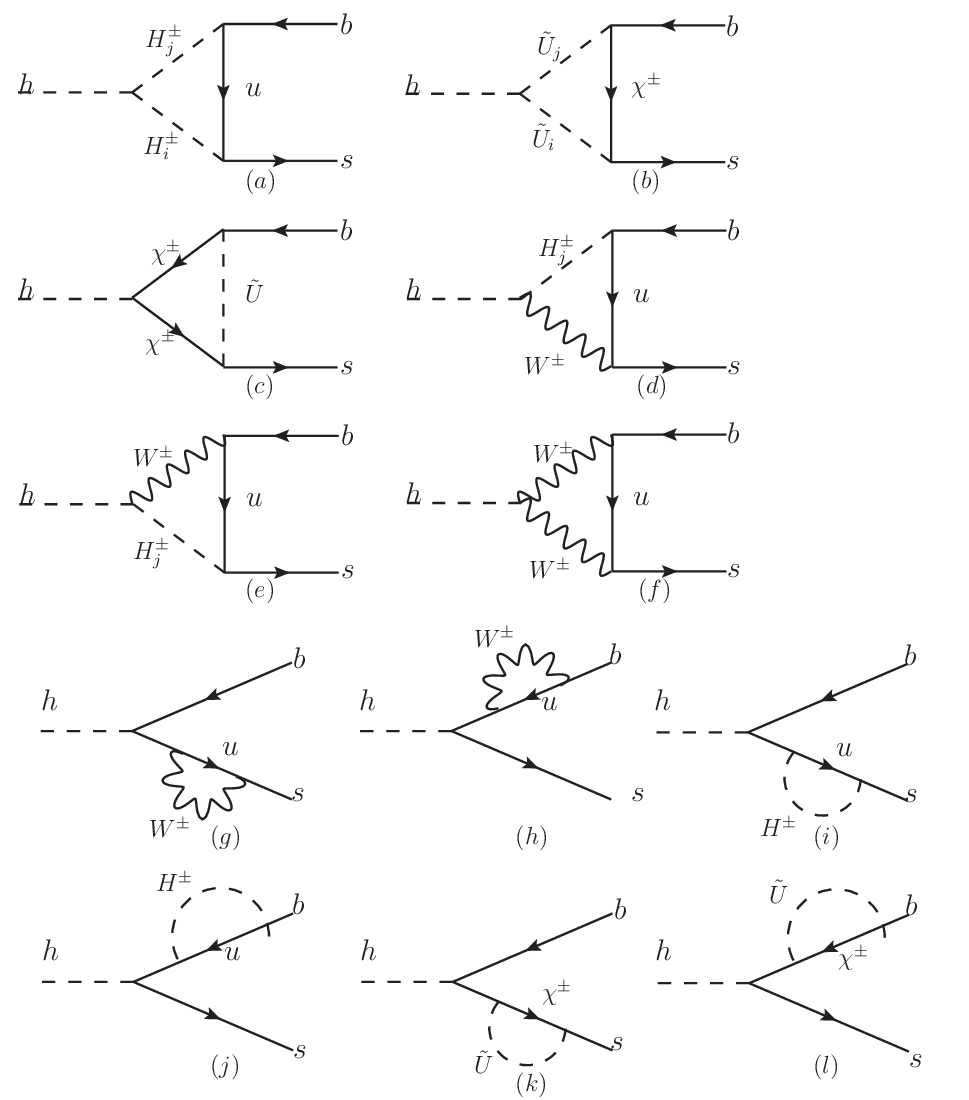}
	\caption[]{Feynman diagrams for the $h \rightarrow bs$ process.}
	\label{h-bs}
\end{figure}

\subsection{The rare decay $h\rightarrow bs$}
 In this section, we investigate the amplitude and branching ratio of $h \rightarrow bs$. The corresponding Feynman diagrams are depicted in Fig.$\ref{h-bs}$. As a specific example, we analyze the diagram in Fig.$\ref{h-bs}$(b), whose amplitude is given by
 \begin{eqnarray}
 	&&i\mathcal{M}=-\bar{s}(p+q)\int\frac{d^Dk}{(2\pi)^D}\frac{1}{[k^2
 		-m_{{\chi}^-_k}^2][(p-k)^2-m_{\tilde{U}_j}^2][(p+q-k)^2-m_{\tilde{U}_i}^2]} \nonumber
 	\end{eqnarray}
 \begin{eqnarray}
 	&&\times\Big( (B_LP_L+B_RP_R)(\not{k} +m_{{\chi}^-_k})
 	(A_LP_L+A_RP_R)\Big)C b(p).\label{x0}
 \end{eqnarray}

In this expression, $p$ represents the momentum of the bottom quark ($b$), $(p+q)$ corresponds to the momentum of the strange quark ($s$), and $k$ is the loop momentum. The $m_{\tilde{U}_{i(j)}}$ term corresponds to the mass of the up-squark. $A_L, B_L, A_R, B_R$ and $C$ represent the coupling vertices.

$A_L$ and $A_R$ are left-handed and right-handed couplings of the vertex $\chi^{-}_kd_i\tilde{U}_j$:
\begin{eqnarray}
	&&A_L = U_{j2}^*\sum_{b=1}^3Z_{kb}^{U,*}\sum_{a=1}^3V_{ia}^{*}Y_{d,ab}\nonumber\\&&
	A_R = (-g_2\sum_{a=1}^3Z_{ka}^{U,*}V_{j1}+\sum_{b=1}^3\sum_{a=1}^3Y_{u,ab}^*Z_{k3+a}^{U,*}V_{j2})V_{ia}.
\end{eqnarray}
$B_L$ and $B_R$ are left-handed and right-handed couplings of the vertex $\bar{d}_j\chi^{-}_k\tilde{U}_i$:
\begin{eqnarray}
	&&B_L = (-g_{2}V^{*}_{i1}\sum_{a=1}^{3}Z^{U}_{ka}+V^{*}_{i2}\sum_{b=1}^{3}\sum_{a=1}^{3}Y_{u,ab}Z^{U}_{k3+a})V^*_{jb}\nonumber\\&&
	B_R = \sum_{b=1}^3\sum_{a=1}^3Y^*_{d,ab}Z^U_{kb}U_{i2}V_{jb}.
\end{eqnarray}
$C$ is the coupling constant of $h\tilde{U}_i \tilde{U}_j$. The subscripts $L$ and $R$ represent the left$-$handed and right$-$handed parts, respectively.

Subsequently, we apply Feynman parametrization together with D-dimensional regularization to perform the integration over the denominator. The ultraviolet divergences are then properly subtracted or regulated, yielding the final expression presented in Eq.~(\ref{x0}).
\begin{eqnarray}
&&	\mathcal{M} =\bar{s}(p+q) \int_{0}^{1} dx \int_{0}^{1} 2y \, dy \Bigl\{
	(A_L B_R C \left[ (1-x) y m_b + (1-y) m_s \right] \nonumber \\&&
	\hspace{1.0cm} + A_L B_L C m_{\chi^-_k})P_L + (L \to R) \Bigr\}
	\times \frac{1}{32\pi^2 T}b(p), \nonumber \\&&
	T = x y m_{\chi^-_k}^2 + (1-x) y m_{\tilde{U}_i}^2 + (1-y) m_{\tilde{U}_j}^2 \nonumber \\&&
	\hspace{1.0cm} + (1-x) y \bigl[ (1-x) y - 1 \bigr] p^2 - (1-y) y (p + q)^2 \nonumber \\&&
	\hspace{1.0cm} + 2 (1-x) y (1-y) \, p \cdot (p + q).
\end{eqnarray}

 The decay width $\Gamma$ of $h \rightarrow bs$ is obtained by substituting $|\mathcal{M}|^2$ into the following formula\cite{Arco:2023hmz}
 \begin{eqnarray}
 	&&\Gamma(h\rightarrow bs) = 2N_c\frac{\lambda^\frac{1}{2}(m^2_h,m^2_b,m^2_s)}{16\pi m^3_h}|\mathcal{M}|^2,
 \end{eqnarray}
 here, $N_c$ = 3 is a color factor, $\lambda(x,y,z) = (x - y - z)^2 - 4y^2z^2$.

 The branching ratio we obtained is
 \begin{eqnarray}
 	&&\mathrm{Br}(h\rightarrow bs) = \frac{\Gamma(h\rightarrow bs)}{\Gamma(h)}.
 \end{eqnarray}
 Here, $\Gamma(h) = 4.1 \times 10^{-3} \rm{GeV}$\cite{LHCHiggsCrossSectionWorkingGroup:2016ypw}. We adopt the theoretical SM-like Higgs total width $\Gamma_h$ for the following reasons. On the one hand, the direct experimental determination of the Higgs total width from ATLAS and CMS is still much less precise and is usually model dependent, as it relies on specific assumptions about the decay channels. On the other hand, using the theoretical value allows a direct comparison with the SM prediction and is more self-consistent in the context of new physics calculations. 

\section{Numerical Results}

In this section, we analyze the numerical results while incorporating the relevant experimental constraints. Given that the experimental bounds from $\bar{B} \to X_s \gamma$ tightly restrict the parameter space of the NB-LSSM, it is essential to account for the impact of $\bar{B} \to X_s \gamma$ when studying the $h \to b s$ process. Below, we discuss in detail the key parameters that are sensitive to both processes. The following restrictions are taken into account in the numerical analysis.
\begin{enumerate}
	\item Considering the updated experimental data on searching \( Z' \) indicates \( M_{Z'} \geq 5.15 \, \text{TeV} \) at 95\% C.L.~\cite{CMS:2022eud}, we choose \( M_{Z'} = 5.2 \, \text{TeV} \) in the follows.
	\item The ratio between the \( Z' \) mass and its gauge coupling at 99\% C.L. as \( M_{Z'}/g_B \geq 6 \, \text{TeV} \)~\cite{GCG,MAB}, and then the scope of \( g_B \) is \( 0 < g_B < 0.86 \).
	\item LHC experimental data constrain \( \tan \beta' < 1.5 \)~\cite{48}.
	\item For particles that exceed the SM, the mass limits are considered: the chargino mass is greater than 1.3 TeV~\cite{navas2024,LEPChargino,ATLAS:2025ew,ATLAS:2025LL}, the charged Higgs mass is greater than 0.6 TeV~\cite{ATLAS:2024hya} and the squark mass is greater than 2 TeV~\cite{Un:2016hji}.
	\item The lightest CP-even Higgs mass is around \( m_h = 125.20 \pm 0.11 \, \text{GeV} \)~\cite{ATLAS:2023oaq}.
	\item The Higgs boson decays $(h \rightarrow \gamma\gamma, ZZ^*, WW^*, b \bar b, \tau \bar\tau)$ should be satisfied~\cite{ATLAS:LFVHiggs2023}.
\end{enumerate}
\begin{table}[!h]
    \centering
    \setlength{\tabcolsep}{20pt}
    \caption{Scanning parameters for Fig.~\ref{fig:combined}}
    \label{tab:scan_params}
    \begin{tabular}{|l|c|c|c|c|c|c|c|}
        \hline
        \textbf{Parameters} & \( v_S / \text{TeV} \) & \( T_{2}, T_{\lambda} / \text{TeV} \) & \( T_{\kappa} / \text{TeV} \) & \(\tan \beta\) & \(\tan \beta'\) \\ \hline
        \textbf{Min}        & 0.5                   & -2                                   & -2                         & 5              & 1              \\ \hline
        \textbf{Max}        & 5                     & 2                                    & 3                          & 60             & 1.5            \\ \hline
    \end{tabular}
    \vspace{3pt}
    \begin{tabular}{|l|c|c|c|c|c|c|}
        \hline
        \textbf{Parameters} & \( g_B \) & \( g_{YB} \) & \(\lambda\) & \(\lambda_2\) & \(\kappa\) & \(\delta _{23}^{LL}, \delta _{23}^{RR}, \delta _{23}^{LR}\) \\ \hline
        \textbf{Min}        & 0.1      & -0.45        & 0.1         & -1            & -1        & \( 1 \times 10^{-3} \) \\ \hline
        \textbf{Max}        & 0.85     & -0.05        & 1.4         & 1             & 3         & \( 1 \) \\ \hline
    \end{tabular}
\end{table}

In the calcaulations, we take the up quark mass $m_u$ = 2.2 MeV, the down quark mass $m_d$ = 4.7 MeV, the strange quark mass $m_s$ = 0.095 GeV, the bottom quark mass $m_b$ = 4.18 GeV, the charm quark mass $m_c$ = 1.275 GeV, the top quark mass $m_t$ = 173.5 GeV, \( m_W = 80.385 \, \text{GeV} \), \( m_Z = 91.188 \,\text{GeV} \), \( \alpha_{\text{em}}(m_Z) = \frac{1}{128.9} \), \( \alpha_s(m_Z) = 0.119 \). The combined searches performed by the ALEPH, DELPHI, L3, and OPAL
Collaborations at LEP2 set a lower limit of
$m_{\tilde{\chi}_1^\pm}>103.5~{\rm GeV}$ at the $95\%$ confidence
level for promptly decaying charginos with heavy sneutrinos and a
sufficiently large chargino--neutralino mass splitting
\cite{LEPChargino}. Recent searches by the ATLAS Collaboration provided constraints on electroweakly produced charged fermions under different assumptions\cite{ATLAS:2025ew}.
A search in final states containing hadronically decaying $\tau$
leptons and $b$-jets has been interpreted in an R-parity violation supersymmetric
scenario, yielding observed lower limits of $880$ GeV and $1170$ GeV
for higgsino-like and wino-like states, respectively. A complementary
search for long-lived charged particles excludes nearly pure-wino
charginos below $1.3$ TeV for lifetimes above $100$ ns, assuming a
chargino--neutralino mass splitting of $160$ MeV\cite{ATLAS:2025LL}. These bounds depend on the electroweakino composition, decay topology,
mass splitting, and lifetime, and therefore do not constitute
model-independent limits on the NB-LSSM chargino sector. But they still give us very useful reference. Motivated by these ATLAS searches, we choose the benchmark soft parameters $M_2=1400~{\rm GeV}$ and explicitly calculate the resulting physical chargino masses.

Firstly, the $125\text{ GeV}$ Higgs boson mass and the Higgs signal strengths $\mu_{\gamma\gamma, WW^*, ZZ^*, b\bar{b}, \tau\bar{\tau}}$ have been incorporated into our analysis. The averaged values of the experimental data are derived from the updated PDG~\cite{navas2024,cms2022,atlas-cms2016,cdf-d02013,atlas2020,atlas2021a,atlas2021b,atlas2019}: $\mu^{exp}_{\gamma\gamma}$ = 1.10 $\pm$ 0.07, $\mu^{exp}_{ZZ}$ = 1.01 $\pm$ 0.07, $\mu^{exp}_{WW}$ = 1.19 $\pm$ 0.12, $\mu^{exp}_{b\bar b}$ = 0.98 $\pm$ 0.12, $\mu^{exp}_{\tau\bar\tau}$ = 1.15 $\pm$ 0.15. A random scan is performed over the parameter ranges in Table~\ref{tab:scan_params} to determine the regions consistent with all experimental constraints.
\begin{figure}[t]
	\centering
		\includegraphics[width=0.3\textwidth]{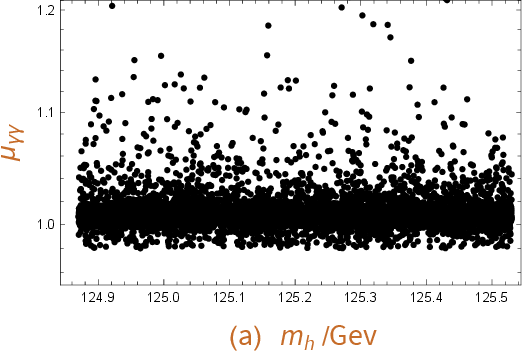}
\hspace{0.5cm}\includegraphics[width=0.3\textwidth]{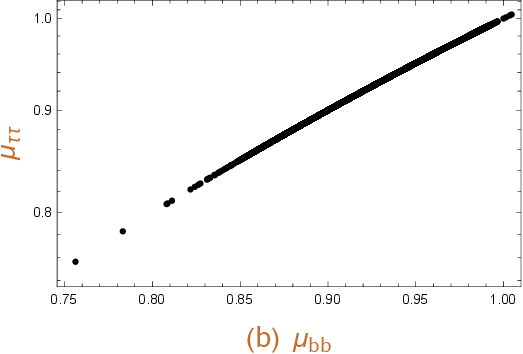}
\hspace{0.5cm}\includegraphics[width=0.3\textwidth]{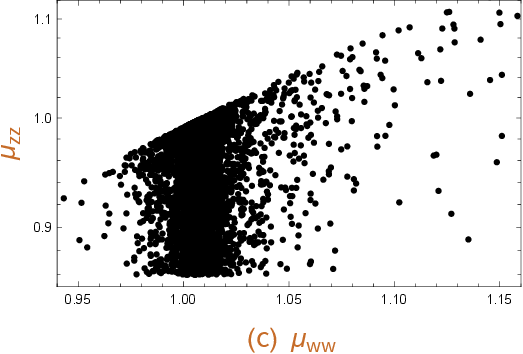}\\
		\includegraphics[width=0.4\textwidth]{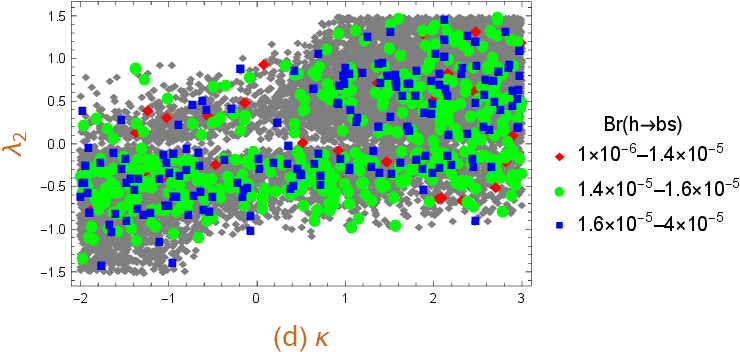}
\hspace{1.5cm}\includegraphics[width=0.4\textwidth]{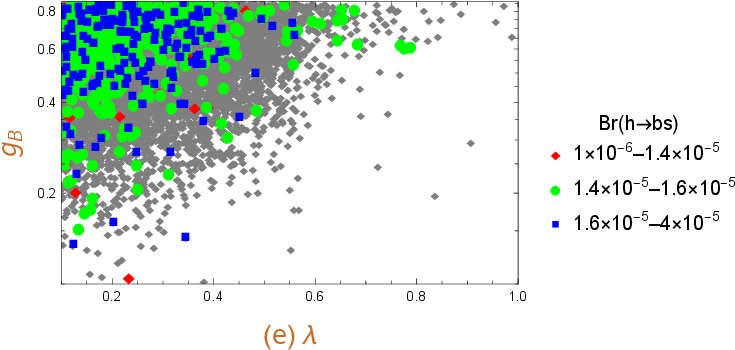}\\
		\includegraphics[width=0.4\textwidth]{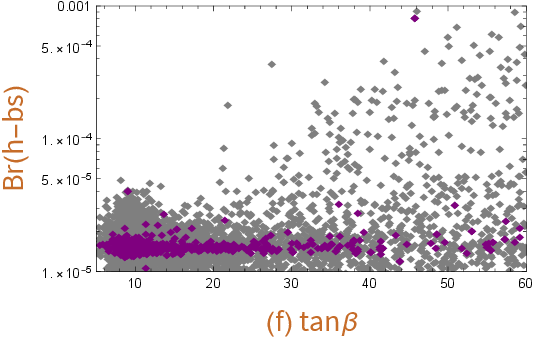}
	\caption[]{ Distributions of key parameters and observables from the scan. All scattered points satisfy the $3\sigma$ experimental constraint on the Higgs mass and the $2\sigma$ constraints on the corresponding Higgs decay signal strengths. Figs.~\ref{fig:combined}(a)-(c) show the Higgs mass and its signal strengths, while (d)-(f) depict the process $h \to bs$ in relation to the key parameters $\kappa$, $\lambda_2$, $\lambda$, $g_B$ and $\tan\beta$. The gray regions represent the parameter space that satisfies the Higgs mass and signal strength constraints. The red, green, blue and purple regions further incorporate the experimental constraints from the $\bar{B} \to X_s \gamma$ process.}
	\label{fig:combined}
\end{figure}

Fig.~\ref{fig:combined}(a) shows the parameter space scan results for the Higgs mass $m_h$ versus the signal strength $\mu_{\gamma\gamma}$. Fig.~\ref{fig:combined}(b) shows a clear positive correlation between $\mu_{\tau\bar{\tau}}$ and $\mu_{b\bar{b}}$ across the entire allowed parameter region, with both quantities varying monotonically in the same direction. This indicates that these two decay channels respond consistently to the new physics contributions in the parameter scan. The majority of points in Fig. 3(c) are concentrated around \(\mu_{WW} \simeq 1.0\), while \(\mu_{ZZ}\) is distributed over a relatively broader interval. In addition, the upper envelope of the distribution gradually shifts toward larger \(\mu_{ZZ}\) values as the \(\mu_{WW}\) increases, and a systematic trend \(\mu_{ZZ} < \mu_{WW}\) is clearly observable for the majority of points. Although both observables share the common dependence on the SM-like Higgs doublet component and the total Higgs width, the difference between \(\mu_{ZZ}\) and \(\mu_{WW}\) can be attributed to their different sensitivities to the extended Higgs and gauge sectors in the NB-LSSM. All scanned points satisfy the $3\sigma$ experimental limit on the Higgs mass and the $2\sigma$ limits on the signal strengths, numerically validating the reasonableness of the parameter space chosen in Table~\ref{tab:scan_params}.

Fig.~\ref{fig:combined}(d), (e) and (f) further present the allowed parameter space for $\kappa$, $\lambda_2$, $\lambda$, $g_B$ and $\tan\beta$. The grey regions represent the parameter space satisfying the Higgs mass and signal strength constraints. The ${\color{red}\blacklozenge}$,
${\color{green}\bullet}$,
and ${\color{blue}\blacksquare}$ denote parameter points that additionally satisfy the experimental constraint from the $\bar{B} \to X_s \gamma$ process, with the branching ratio of $h \to bs$ falling within the ranges of $1\times10^{-6}$--$1.4\times10^{-5}$, $1.4\times10^{-5}$--$1.6\times10^{-5}$, and $1.6\times10^{-5}$--$4\times10^{-5}$, respectively. A significant reduction in the number of colour-coded points compared to the grey region is observed, demonstrating that the $\bar{B} \to X_s \gamma$ process imposes a very stringent constraint on the parameter space of the model. In particular, as can be seen from the gray region in panel (f), the branching ratio of $h \to b s$ could in principle reach the order of $10^{-3}$. However, once the constraint from $\bar{B} \to X_s \gamma$ is incorporated, the allowed region for $h \to b s$ is substantially restricted, pushing its branching ratio down to the order of $10^{-5}$.

To understand the dependence of $\mathrm{Br}(h \to bs)$ on the model parameters, The behavior shown in Figs. 3(d)--3(f) can be understood from the structure of the loop amplitudes in Fig. 2. As shown in Fig. 3(d), a clear correlation exists between the parameters $\kappa$ and $\lambda_2$, indicating that they are coupled rather than varying independently within the experimentally allowed parameter space. Specifically, when $\kappa > -1$, $\lambda_2$ typically lies in the range $-1$ to $1.5$; whereas for $\kappa < -1$, $\lambda_2$ tends to be negative. The green points, which represent the predicted branching ratios of $h \to b s$, exhibit a densely clustered distribution. This pattern arises primarily from the stringent constraint imposed by the $\bar{B} \to X_s \gamma$ process. This constraint significantly compresses the allowed parameter region for $h \to b s$, confining its branching ratio to a relatively narrow range across viable parameter combinations, thereby resulting in the observed concentration of green points. The parameters $\kappa$ and $\lambda_2$ mainly exist in the CP-even Higgs mass matrix, which modifies the Higgs rotation matrix $Z^H$ and hence the Higgs-related couplings. Therefore, the $\kappa-\lambda_2$ correlation in Fig. 3(d) is mainly induced by the Higgs boson and signal strength constraints.

Fig.~\ref{fig:combined}(e) displays the distribution of the parameters $\lambda$ and $g_B$ under the experimental constraint from $\bar{B} \to X_s \gamma$. The allowed parameter points are predominantly concentrated in the upper-left region, showing a nonlinear dependence between $\lambda$ and $g_B$. It is noteworthy that
the larger branching ratios for $h \to bs$ are more significantly clustered within the parameter space defined by $\lambda < 0.5$ and $g_B > 0.4$. Consequently, prioritizing this specific region in the model parameter selection facilitates the achievement of a larger $h \to bs$ branching ratio. The parameters \(g_B\) and \(\lambda\) have a more direct impact on the chargino/up-squark diagrams. The gauge coupling \(g_B\) enters the diagonal entries of the up-squark mass matrix through the \(U(1)_{B-L}\) D-term contributions. Consequently, varying \(g_B\) changes the up-squark mass eigenvalues and the rotation matrix \(Z^U\), which in turn modify both the loop denominators and the up-squark-related couplings. Besides, \(\lambda\) affects the chargino mass matrix through the \(\lambda v_S\) term and also contributes to the left-right blocks of the up-squark mass matrix. Therefore, \(\lambda\) controls the chargino/up-squark loop contribution by modifying the chargino masses, the unitary matrices \( U \) and \( V \), the up-squark masses, and the rotation matrix \(Z^U\). The enhancement of \({\rm Br}(h \to bs)\) in the region with relatively small \(\lambda\) and large \(g_B\) is thus caused by the combined effect of modified chargino/up-squark spectra and couplings, and the experimental restriction from \({\rm Br}(\bar B \to X_s \gamma)\).

Fig.~\ref{fig:combined}(f) illustrates the dependence of the $\mathrm{Br}(h \to bs)$ on $\tan\beta$. In the studied parameter space, $\tan\beta$ satisfies all experimental constraints over a wide range. However, the number of viable parameter points that simultaneously fulfill theoretical consistency and experimental bounds is significantly larger within the interval $5 \lesssim \tan\beta \lesssim 40$. This indicates that this region represents a more natural and statistically preferred subspace in the model's parameter space. The parameter $\tan\beta$ determines the values of $v_u$ and $v_d$, and consequently modifies the quark, chargino, up-squark, CP-even/charge Higgs mass matrices and corresponding couplings. Through these modifications, $\tan\beta$ changes both the masses of the particles propagating in the loop diagrams of Fig. 2 and the associated interaction vertices. After imposing the experimental constraint from \({\rm Br}(\bar B \to X_s \gamma)\), most parameter points yielding a strong enhancement are excluded, and the surviving predictions for \({\rm Br}(h \to bs)\) are restricted to a much narrower region, predominantly of order $10^{-5}$. 

\begin{figure}[t]
	\centering
\includegraphics[width=0.35\textwidth]{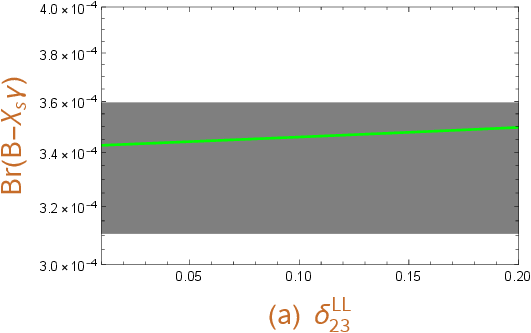}
\hspace{0.5cm}\includegraphics[width=0.35\textwidth]{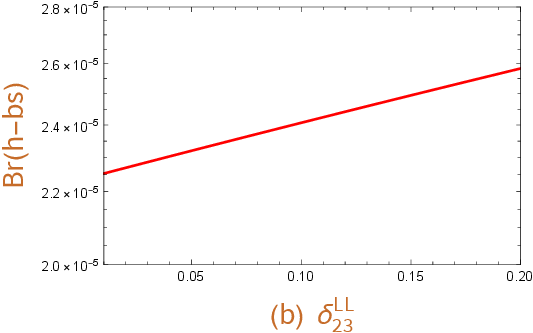}\\
\includegraphics[width=0.35\textwidth]{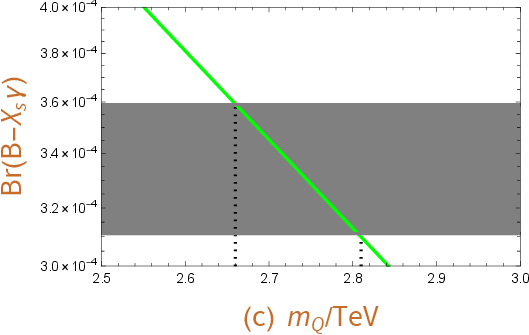}
\hspace{0.5cm}\includegraphics[width=0.35\textwidth]{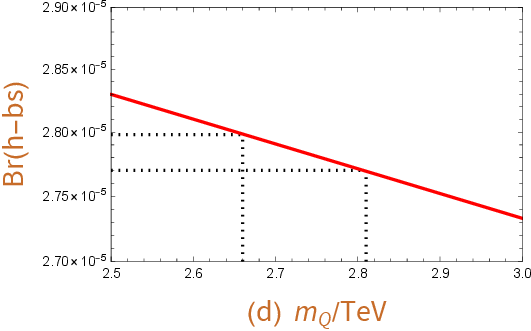}\\
\includegraphics[width=0.35\textwidth]{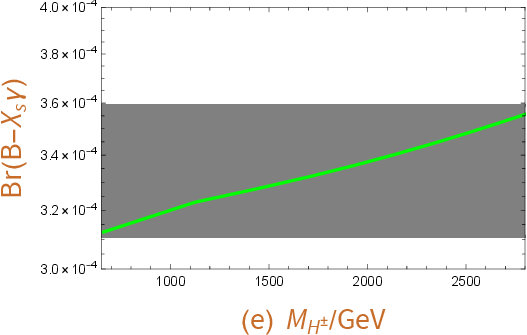}
\hspace{0.5cm}\includegraphics[width=0.35\textwidth]{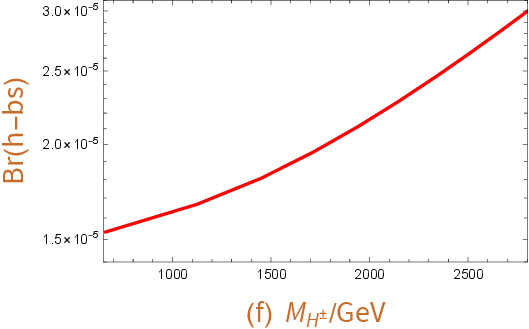}\\
\includegraphics[width=0.35\textwidth]{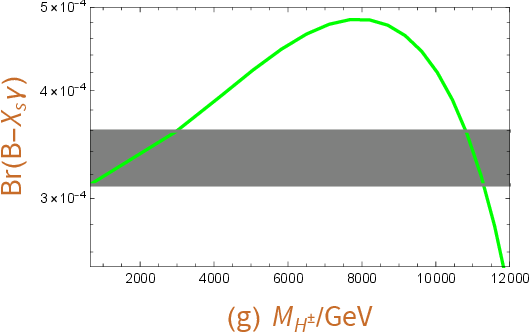}
\hspace{0.5cm}\includegraphics[width=0.35\textwidth]{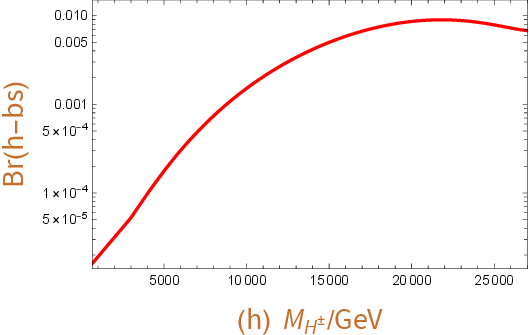}
	\caption[]{Parameter scans for the rare decay processes. The left panels show the branching ratios of $\bar{B} \to X_s \gamma$ (green solid curves), while the right panels display the branching ratios of $h \to bs$ (red solid curves). The gray bands denote the experimental $1\sigma$ intervals for $\bar{B} \to X_s \gamma$, and the black dashed curves show the reachable $h \to bs$ branching ratios under the $\bar{B} \to X_s \gamma$ constraints. From top to bottom: (a)-(b) $\delta _{23}^{LL} $, (c)-(d) $m_{Q}$, and (e)-(h) $M _{H^{\pm}} $.}
	\label{fig:lfv-scans}
\end{figure}

\begin{figure}[t]
	\centering
\includegraphics[width=0.45\textwidth]{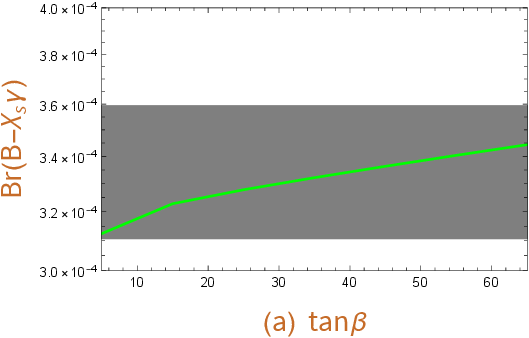}
\hspace{0.5cm}\includegraphics[width=0.45\textwidth]{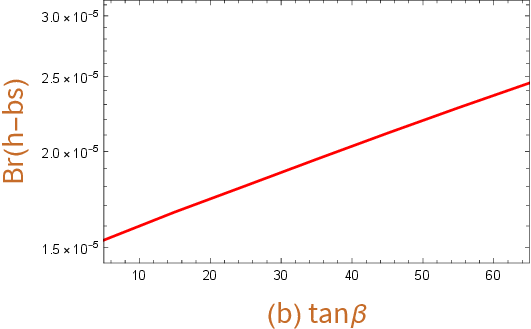}\\
\includegraphics[width=0.45\textwidth]{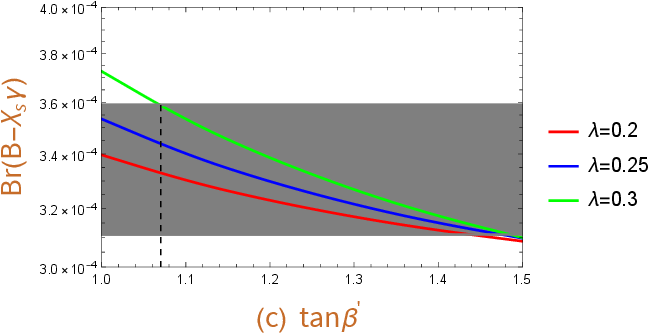}
\hspace{0.5cm}\includegraphics[width=0.45\textwidth]{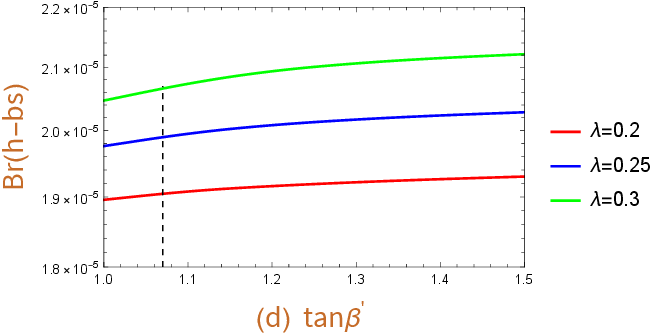}\\
\includegraphics[width=0.45\textwidth]{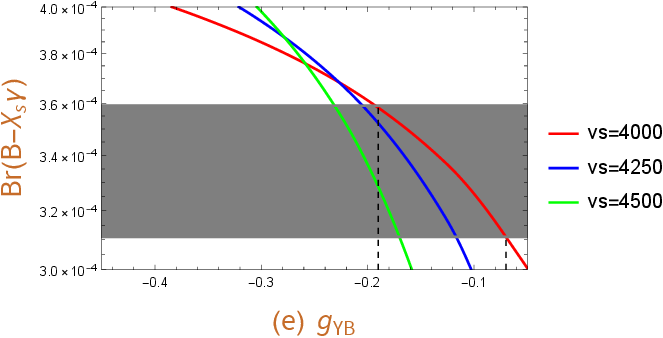}
\hspace{0.5cm}\includegraphics[width=0.45\textwidth]{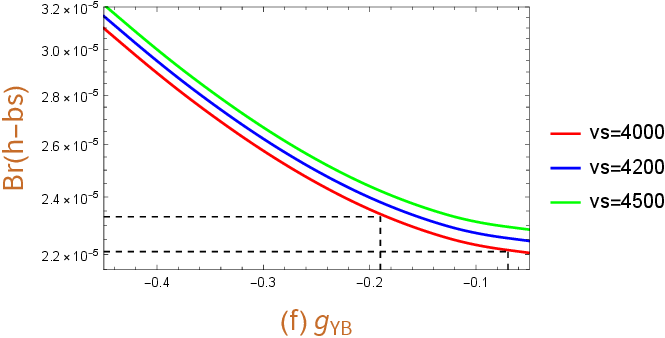}
	\caption{Dependence of the branching ratios for the rare decay processes $\bar{B} \to X_s \gamma$ and $h \to bs$ on key model parameters.
		(a), (c) and (e) show the branching ratio $\mathrm{Br}(\bar{B} \to X_s \gamma)$ as a function of the parameter $\tan\beta$, $\tan\beta'$ and the gauge coupling $g_{YB}$, respectively. In (c)((e)), the dependence of $\mathrm{Br}(\bar{B} \to X_s \gamma)$ on $\tan\beta'$($g_{YB}$) is shown for different values of the coupling parameter $\lambda$(the singlet vacuum expectation value $v_S$). (b), (d), and (f) display $\mathrm{Br}(h \to bs)$ versus $\tan\beta$, $\tan\beta'$ and the gauge coupling $g_{YB}$, respectively. All calculations satisfy the relevant experimental constraints.}
	\label{fig:lfv-scans2}
\end{figure}

A series of one-dimensional scans are presented in Fig.~\ref{fig:lfv-scans} and \ref{fig:lfv-scans2} to show the dependence of the observables on each free parameter. These scans are performed under the combined $3\sigma$ constraint on the Higgs mass and the $2\sigma$ constraints on the Higgs decay channels, with the input parameters set as follows:
\begin{eqnarray}
&&T_2 = 2 \ \mathrm{TeV}, \;
		T_\lambda = -0.76 \ \mathrm{TeV}, \;
		T_{\kappa} = 2.3 \ \mathrm{TeV}, \;
		\delta_{23}^{RR} = 0.6, \;
		\delta_{23}^{LR} = 0.7,\; m_{U}=2.5\ \mathrm{TeV} ,\nonumber\\
&&
		m_{uu}=2.4\ \mathrm{TeV},\; m_{qq}=2.5\ \mathrm{TeV},\;\delta_{12}^{LL} =\delta_{12}^{RR}=\delta_{12}^{LR}=0, \;
		\delta_{13}^{LL} =\delta_{13}^{RR}=\delta_{13}^{LR}=0.
\end{eqnarray}

Fig.~\ref{fig:lfv-scans} displays the behavior of the branching ratios for $\bar{B} \to X_s \gamma$ and $h \to bs$ under one-dimensional parameter scans. As shown in Fig.~\ref{fig:lfv-scans}(a) and (b), we analyze the impact of the off-diagonal mixing parameter $\delta_{23}^{LL}$ in the up-squark mass matrix on the branching ratios of $\bar{B} \to X_s \gamma$ and $h \to bs$. As $\delta_{23}^{LL}$ increases from 0 to 0.20, the branching ratio of $\bar{B} \to X_s \gamma$ exhibits only a minor variation, following an approximately linear and weak growth trend. The branching ratio of $h \to bs$ rises from $2.2\times10^{-5}$ to $2.6\times10^{-5}$, indicating a higher sensitivity of this process to the $\delta_{23}^{LL}$ parameter. This discrepancy demonstrates a clear process dependence in the response of different low-energy observables to the off-diagonal structure of the up-squark sector under the same flavor-violating mixing parameter. Such behavior originate from differences in the coupling structures or propagator contributions of the relevant Wilson coefficients to supersymmetric particle mixing in the respective processes.

Fig.~\ref{fig:lfv-scans}(c) and (d) illustrate the impact of the diagonal element $m_{Q}$ of the up-squark mass matrix: as $m_{Q}$ increases, both branching ratios decrease accordingly. The experimental constraint from $\mathrm{Br}(\bar{B} \to X_s \gamma)$ restricts $m_{Q}$ to a narrow window of 2.66--2.81 TeV. Within this range, the variation in $\mathrm{Br}(h \to bs)$ only decreases slightly from $2.79\times10^{-5}$ to $2.77\times10^{-5}$, indicating a very weak dependence that has effectively reached saturation. Moreover, the diagonal element $m_Q$ suppresses the branching ratios through its characteristic $\frac{1}{m_Q^2}$ scaling, whereas the off-diagonal mixing parameter $\delta_{23}^{LL}$ provides a direct enhancement, especially for $h \to bs$, by introducing new flavor-violating couplings. This opposite role of diagonal and off-diagonal parameters clearly reflects the distinct mechanisms through which the supersymmetric flavor structure affects different low-energy observables.

Figs.~\ref{fig:lfv-scans}(e)-(h) depict the dependence of the $\mathrm{Br}(\bar{B} \to X_s \gamma)$ and $\mathrm{Br}(h \to bs)$ on the charged Higgs mass $M_{H^\pm}$. As shown in Fig.~\ref{fig:lfv-scans}(e), within the parameter space allowed by the $\bar{B} \to X_s \gamma$ constraint (corresponding to $M_{H^\pm}$ in the range of approximately 600--2800 GeV), $\mathrm{Br}(\bar{B} \to X_s \gamma)$ increases with $M_{H^\pm}$. Simultaneously, Fig.~\ref{fig:lfv-scans}(f) reveals that $\mathrm{Br}(h \to bs)$ also exhibits a significant growing trend across this mass window, reaching a value up to $3 \times 10^{-5}$. This indicates a consistent dependence of both decay processes on $M_{H^\pm}$ within this specific mass range. The decoupling behavior of large $M_{H^\pm}$ shown in Fig.~4(g) and (h) are discussed in Appendix C.

Fig.~\ref{fig:lfv-scans2}(a) and (b) illustrate the dependence of the branching ratios for $\bar{B} \to X_s \gamma$ and $h \to bs$ on the parameter $\tan\beta$. Within the parameter space consistent with the experimental constraints at the $1\sigma$ level, the branching ratio of $\bar{B} \to X_s \gamma$ exhibits a growing trend as $\tan\beta$ increases from 5 to 65. Meanwhile, the branching ratio of $h \to bs$ rises monotonically from $1.5 \times 10^{-5}$ to $2.5 \times 10^{-5}$. This pronounced sensitivity of both decay processes to $\tan\beta$ clearly identifies it as a key governing parameter for these flavor-changing decays.

A small kink is observed in Fig.~5(a) around $\tan\beta \approx 15$ in the $\mathrm{Br}(\bar{B} \to X_s \gamma)$ curve. This feature originates from a crossover between the two classes of dominant one-loop contributions to the effective Wilson coefficient associated with the four diagrams in Fig.~1. One class corresponds to the charged-Higgs loop contributions (Fig.~1(a) and~(b)), which decrease in the interval $\tan\beta<15$.
This behavior may be caused mainly by the rapid suppression of the
$m_t\cot\beta$ component of the $H^\pm\bar t b$ coupling, together with
the increase of the charged-Higgs mass and the corresponding
decoupling of the charged-Higgs loop functions. And then the charged-Higgs loop contributions (Fig.~1(a) and~(b)) increase as $\tan\beta>15$, which may be caused mainly by the rapid increase of the $m_b\tan\beta$ component of the $H^\pm\bar t b$ coupling, although still together with the increase of the charged-Higgs mass and the corresponding decoupling of the charged-Higgs loop functions. The other class corresponds to the chargino--squark loop contributions (Fig.~1(c) and~(d)), which increase linear as $\tan\beta$ enlarges and becomes dominant for $\tan\beta>15$.  Then, the total $\mathrm{Br}(\bar{B} \to X_s \gamma)$ combining the contributions from these two classes possesses a noticeably faster growth as $\tan\beta>15$ than $\tan\beta<15$. The kink therefore reflects a change in the dominant loop contribution rather than a numerical discontinuity or a new kinematic threshold.

Fig.~\ref{fig:lfv-scans2}(c) and (d) show the variation of the branching ratios with $\tan\beta'$ for different values of $\lambda$. From Fig.~\ref{fig:lfv-scans2}(c), it can be seen clearly that when $\lambda = 0.3$, $\mathrm{Br}(\bar{B} \to X_s \gamma)$ reaches a relatively large value, with the allowed parameter range for $\tan\beta'$ being 1.07--1.5. Within this range, the branching ratio gradually decreases as $\tan\beta'$ increases. In addition, a smaller $\lambda$ also allows for a wider viable parameter space under the experimental constraints. For a fixed $\tan\beta'$, $\mathrm{Br}(\bar{B} \to X_s \gamma)$ increases with $\lambda$. Similarly, Fig.~\ref{fig:lfv-scans2}(d) indicates that $\mathrm{Br}(h \to bs)$ grows with $\lambda$ at fixed $\tan\beta'$, while $\tan\beta'$ itself has a rather weak influence on this decay channel. At $\lambda = 0.3$, a slight increase of $\mathrm{Br}(h \to bs)$ with $\tan\beta'$ can be clearly observed.

Finally, Fig.~\ref{fig:lfv-scans2}(e) and (f) investigate the dependence of $\mathrm{Br}(\bar{B} \to X_s \gamma)$ and $\mathrm{Br}(h \to bs)$ on the gauge coupling $g_{YB}$ for different values of the vacuum expectation value $v_S$. As shown in Fig.~\ref{fig:lfv-scans2}(e), for a fixed $v_S$, $\mathrm{Br}(\bar{B} \to X_s \gamma)$ decreases significantly with increasing $g_{YB}$ (from $-0.45$ to $-0.15$), indicating a suppressing role of $g_{YB}$. Meanwhile, for a given $g_{YB}$, a smaller $v_S$ leads to a slightly larger viable parameter space and yields a larger branching ratio. The three curves are clearly separated, reflecting the reduction effect of $v_S$. Fig.~\ref{fig:lfv-scans2}(f) shows that $\mathrm{Br}(h \to bs)$ exhibits qualitatively similar behavior: it also decreases rapidly with increasing $g_{YB}$. However, for a fixed $g_{YB}$, a larger $v_S$ results in a larger branching ratio. Thus, while both decays are suppressed by larger $g_{YB}$, the role of $v_S$ is opposite for the two processes. Differences such as the rate of decrease with $g_{YB}$ and the degree of separation between curves corresponding to different $v_S$ values arise because these parameters carry different weights or interference effects in the amplitude of each specific process.

\section{discussion and conclusion}
In this work, we have systematically investigated the decay behaviors of $h \to bs$ and $\bar{B} \to X_s \gamma$ processes within the NB-LSSM framework. Through a global parameter scan, incorporating the $3\sigma$ experimental limits on the Higgs mass, $2\sigma$ signal strength constraints from Higgs decays, and $1\sigma$ observational bounds on $\bar{B} \to X_s \gamma$, we have identified the viable regions of the physical parameter space. Our results indicate that, without the constraints from $\bar{B} \to X_s \gamma$, the branching ratio of $h \to bs$ can reach up to $\mathcal{O}(10^{-3})$, whereas it is suppressed to only about $10^{-5}$ once these constraints are included. Hence, the experimental data on $\bar{B} \to X_s \gamma$ impose a strong constraint on the $h \to bs$ decay channel.

Further one-dimensional parameter scans reveal the regulatory roles of key physical parameters in these decay processes. The off-diagonal and diagonal elements of the supersymmetric flavor structure play distinct roles. The off-diagonal squark mixing parameter $\delta_{23}^{LL}$ significantly enhances the branching ratio for $h \to bs$ while having a negligible impact on $\bar{B} \to X_s \gamma$, revealing a clear process dependence in how low-energy observables respond to flavor violation. Conversely, the diagonal squark mass parameter $m_Q$ suppresses both branching ratios via a characteristic $\frac{1}{m_Q^2}$ scaling, with its value tightly constrained to the narrow range of approximately 2.66 to 2.81 TeV by the experimental bounds on $\bar{B} \to X_s \gamma$. The study identifies the charged Higgs mass $M_{H^\pm}$ and $\tan\beta$ as key governing parameters common to both processes. Within the experimentally allowed mass window (approximately 600--2800 GeV), both branching ratios increase with $M_{H^\pm}$. Simultaneously, an increase in $\tan\beta$ (from 5 to 65) monotonically raises both branching ratios, highlighting its central role in regulating such flavor-changing decays.

Furthermore, the analysis of other model parameters such as $\tan\beta'$, $\lambda$, $g_{YB}$, and $v_S$ reveals their influence on the decay processes. When  $\tan\beta'$ is fixed, both branching ratios increase significantly with increasing $\lambda$. However, $\tan\beta'$ itself has only a relatively weak effect on the branching ratio of $h \to bs$. The gauge coupling $g_{YB}$ exhibits a significant suppressing effect on both processes. However, the role of the vacuum expectation value $v_S$ is opposite for the two processes: for $\bar{B} \to X_s \gamma$, a smaller $v_S$ leads to a larger branching ratio, whereas for $h \to bs$, a larger $v_S$ yields a larger branching ratio. These differences stem from the varying weights and interference effects of these parameters within the amplitude of each specific process. This work elucidates the differential impacts of various parameters on flavor-changing processes by comparing the responses of two low-energy flavor-changing processes to supersymmetric model parameters. It not only provides important constraints on the parameter space of new physics beyond the Standard Model but also establishes a theoretical foundation for future experimental searches for rare Higgs decay channels.

{\bf Acknowledgments}
This work is supported by the Major Project of National Natural Science Foundation of China (NNSFC) (No. 12235008), the National Natural Science Foundation of China (NNSFC) (No. 12075074, No. 12075073), the Natural Science Foundation of Hebei province(No.A2022201022, No. A2023201041), the Natural Science Foundation of Hebei Education Department(No. QN2022173), the Project of the China Scholarship Council (CSC) (No. 202408130113). This work is also supported by Funda\c{c}\~{a}o para a Ci\^{e}ncia e a Tecnologia (FCT, Portugal) through the projects UID/00777/2025 (https://doi.org/10.54499/UID/00777/2025).

\appendix
\section{The Wilson coefficients of the $B\rightarrow X_s \gamma$ process }\label{bsr}
\vspace{-1cm}
\begin{eqnarray}
&&		\hspace{-1cm}C_{7,NP}^{(a)}(\mu_{EW}) =\frac{-v^2}{m_b V^*_{ts} V_{tb}}
		\Bigg\{
		\frac{1}{2\Lambda^2} \left[ I_3( x_{u_j}, x_{H_i^{\pm}} ) - I_4( x_{u_j}, x_{H_i^{\pm}})  \right]
		\bigl( m_b C_{H_i^{\pm} s u_j}^L C_{H_i^{\pm} b \bar{u}_j}^R
		+ m_{s} C_{H_i^{\pm} s u_j}^R C_{H_i^{\pm} b \bar{u}_j}^L \bigr) \nonumber\\
&&\hspace{1.4cm}\quad + \frac{1}{\Lambda^2} m_{u} \left[ I_1( x_{u_j}, x_{H_i^{\pm}}) - I_3( x_{u_j}, x_{H_i^{\pm}} ) \right]
		C_{H_i^{\pm} s u_j}^L C_{H_i^{\pm} b \bar{u}_j}^L
		\Bigg\},\nonumber\\
&&\hspace{-1cm}
		C_{7,NP}^{(b)}(\mu_{EW}) = \frac{2 v^2}{3V^*_{ts} V_{tb}m_b}
		\Bigg\{
		\frac{1}{2\Lambda^2} \left[ I_3( x_{H_i^{\pm}}, x_{u_j} )  - I_4( x_{H_i^{\pm}}, x_{u_j} )\right]
		(m_b C_{H_i^{\pm} s u_j}^L C_{H_i^{\pm} b \bar{u}_j}^R + m_{s} C_{H_i^{\pm} s u_j}^R C_{H_i^{\pm} b \bar{u}_j}^L ) \nonumber\\
&&\hspace{1.4cm}\quad + \frac{1}{\Lambda^2} m_{u} I_3( x_{H_i^{\pm}}, x_{u_j} )
		C_{H_i^{\pm} s u_j}^L C_{H_i^{\pm} b \bar{u}_j}^L
		\Bigg\},\nonumber\\
&&\hspace{-1cm}
		C_{7,NP}^{(c)}(\mu_{EW}) =\frac{-2v^2}{3 m_b V^*_{ts} V_{tb}}
		\Bigg\{
		\frac{1}{2\Lambda^2} \left[ I_3( x_{\chi_j^{\pm}}, x_{\tilde{U}_i} ) - I_4(x_{\chi_j^{\pm}}, x_{\tilde{U}_i})  \right]
		\bigl( m_b C_{\tilde{U}_i s \chi_j^{\pm}}^L C_{\tilde{U}_i b \chi_j^{\pm}}^R
		+ m_{s} C_{\tilde{U}_i s \chi_j^{\pm}}^R C_{\tilde{U}_i b \chi_j^{\pm}}^L \bigr) \nonumber\\
&&\hspace{1.4cm}\quad + \frac{1}{\Lambda^2} m_{\chi^\pm_j} \left[ I_1(x_{\chi_j^{\pm}}, x_{\tilde{U}_i}) - I_3( x_{\chi_j^{\pm}}, x_{\tilde{U}_i}) \right]
		C_{\tilde{U}_i s \chi_j^{\pm}}^L C_{\tilde{U}_i b \chi_j^{\pm}}^L
		\Bigg\},\nonumber\\
&&\hspace{-1cm}
		C_{7,NP}^{(d)}(\mu_{EW}) = \frac{v^2}{m_b V^*_{ts} V_{tb}}
		\Bigg\{
		\frac{1}{2\Lambda^2} \left[ I_3( x_{\tilde{U}_i}, x_{\chi_j^{\pm}} ) - I_4( x_{\tilde{U}_i}, x_{\chi_j^{\pm}} )\right]
		(m_b C_{\tilde{U}_i s \chi_j^{\pm}}^L C_{\tilde{U}_i b \chi_j^{\pm}}^R + m_{s} C_{\tilde{U}_i s \chi_j^{\pm}}^R C_{\tilde{U}_i b \chi_j^{\pm}}^L ) \nonumber\\
&&\hspace{1.4cm}\quad + \frac{1}{\Lambda^2} m_{\chi^\pm_j} I_3( x_{\tilde{U}_i}, x_{\chi_j^{\pm}} )
		C_{\tilde{U}_i s \chi_j^{\pm}}^L C_{\tilde{U}_i b \chi_j^{\pm}}^L
		\Bigg\},\nonumber\\
&&\hspace{-1cm}
		C_{7,NP}^{'(n)}(\mu_{EW}) = C_{7,NP}^{(n)}(\mu_{EW})(L \leftrightarrow R), \quad (n = a, b, c, d).
\end{eqnarray}

\section{One-loop functions} \label{OLF}
In this section, we give out the corresponding one-loop integral functions, which read as:
\begin{eqnarray}
&&\hspace{-1cm}	I_1(x,y) =  \frac{\log y + 1}{y - x} + \frac{x \log x - y \log y}{(y - x)^2} , \nonumber\\
&&\hspace{-1cm}
	I_3(x,y) = \frac{1}{2 } ( \frac{2 \log y + 3}{y - x} - \frac{2 y + 4 y \log y}{(y - x)^2}  - \frac{2 x^2 \log x + 2 y^2 \log y}{(y - x)^3}),
\nonumber\\
&&\hspace{-1cm}
	I_4(x,y) = \frac{1}{6 } ( \frac{6 x^3 \log x - 6 y^3 \log y}{(y - x)^4} + \frac{6 y^2 + 18 y^2 \log y}{(y - x)^3} + \frac{6 \log y + 11}{y - x} - \frac{15 y + 18 y \log y}{(y - x)^2} ).
\end{eqnarray}
\section{The decoupling behavior of increasing $M_{H^\pm}$} 
We agree that, if all couplings and the remaining particle spectrums are kept fixed,
the contribution of an isolated charged-Higgs loop would decrease with
increasing $M_{H^\pm}$ because the corresponding loop functions exhibit
the usual decoupling behavior.

Actually, the dependence displayed in Fig.~4(f) is not obtained by varying $M_{H^\pm}$ as an independent parameter while keeping the other
quantities fixed. In our numerical analysis, the charged-Higgs mass is
a derived quantity whose increase is mainly induced by varying
$\tan\beta$. Therefore, the horizontal axis in Fig.~4(f) also implicitly
represents a correlated variation of $\tan\beta$. Since $\tan\beta=v_u/v_d$ determines the relative sizes of the two Higgs-doublet vacuum expectation values, its variation modifies not only
the charged-Higgs mass, but also the quark Yukawa couplings, the
chargino and up-squark mass matrices, the CP-even Higgs mixing matrix,
and the corresponding interaction vertices. Therefore, it shows the total branching ratio of \(h \to bs\) after summing all diagrams in Fig. 2 and after imposing the constraint from \({\rm Br}(\bar B \to X_s \gamma)\).

Taking Fig. 2(a) and (b) as an example, the charged-Higgs part of the amplitude contains charged-Higgs coupling to quarks and loop function involving \(M_{H^\pm}\).
The loop function is suppressed for large \(M_{H^\pm}\) if the couplings are fixed. However, the charged-Higgs coupling to quarks contains the chiral structures
\begin{eqnarray}
H^+\bar u d\propto
m_u\cot\beta\,P_L+m_d\tan\beta\,P_R,
\end{eqnarray}
which correlates the parameter $\tan\beta$ directly. In the allowed mass window shown in Fig. 4(f), the simultaneous
$\tan\beta$-induced changes in this coupling/interference effect may overcome the propagator suppression, leading to an increasing total \({\rm Br}(h \to bs)\) in a certain parameter interval. This does not contradict the usual decoupling behavior in the asymptotic fixed-coupling limit. The decoupling behavior can be reflected by the new added Fig. 4(g) and (h).

We also emphasize that this increasing behavior occurs only within the
phenomenologically relevant finite interval shown in Fig.~4(f). A
further increase of $M_{H^\pm}$ in this particular one-dimensional
trajectory would require very large values of $\tan\beta$. Such points
are strongly constrained, or excluded, by the measured
${\rm Br}(\bar B\to X_s\gamma)$, the Higgs-sector observables, and the
relevant charged-Higgs searches. They are therefore not part of the
viable parameter space considered in our phenomenological analysis.

Accordingly, Fig.~4(f) should be interpreted as showing a correlated
finite-range dependence along the allowed parameter trajectory, rather
than the asymptotic decoupling behavior obtained by varying
$M_{H^\pm}$ independently.

\end{document}